\documentclass[12pt]{JHEP3}

\usepackage{amssymb, amsmath, amsopn, amsthm}

\usepackage{epsfig}

\usepackage{graphicx}

\def\makeatletter{\catcode`\@=11}
\makeatletter
\def\mathbox#1{\hbox{$\m@th#1$}}%
\def\math@ccstyles#1#2#3#4#5#6#7{{\leavevmode
      \setbox0\mathbox{#6#7}%
      \setbox2\mathbox{#4#5}%
      \dimen@ #3%
      \baselineskip\z@\lineskiplimit#1\lineskip\z@
      \vbox{\ialign{##\crcr
             \hfil \kern #2\box2 \hfil\crcr
             \noalign{\kern\dimen@}%
             \hfil\box0\hfil\crcr}}}}
\def\mathaccstyles{\math@ccstyles\maxdimen}
\def\maththroughstyles{\math@ccstyles{-\maxdimen}}
\def\unity%
 {\maththroughstyles{.45\ht0}\z@\displaystyle {\mathchar"006C}\displaystyle 1}

\title{ Entangled  Dilaton Dyons}

\author{Nilay Kundu$^1$, Prithvi Narayan$^1$,  Nilanjan Sircar$^1$ and Sandip P. Trivedi$^1$

 ~\\

$^1$Tata Institute for Fundamental Research  \\
Mumbai 400005, India\\

\vspace{0.1cm}

\email{Email: nilay.tifr@gmail.com, prithvi.narayan@gmail.com,  nilanjan.tifr@gmail.com, trivedi.sp@gmail.com} \\

}

\abstract{  Einstein-Maxwell theory coupled to a dilaton is known to give rise to extremal solutions with  hyperscaling
violation. We study the behaviour of these solutions in the presence of a small magnetic field. We  find that in a  region
of  parameter space the magnetic field is relevant in the infra-red and completely changes the behaviour of the solution
which now flows to an  $AdS_2\times R^2$ attractor.  As a result there is an extensive  ground state entropy
and the entanglement entropy  of a sufficiently big region on the boundary grows like the volume. In particular,
this happens for values of parameters at  which the purely electric theory has an entanglement entropy  growing with the area, $A$,  like $A \log(A)$ which
is believed to be a characteristic feature of a Fermi surface.
 Some other thermodynamic properties  are also analysed and a more detailed characterisation of the entanglement entropy is also carried out in the presence of a magnetic field.  Other regions of parameter
space not described by the $AdS_2\times R^2$ end point are  also discussed.}

\preprint{
{\normalsize TIFR/TH/12-32} \\
}

\def\be{\begin{equation}}

\def\ee{\end{equation}}

\def\bea{\begin{eqnarray}}

\def\eea{\end{eqnarray}}

\begin{document}


\section{Introduction}

The AdS/CFT correspondence suggests  that    interesting   connections could arise between  gravitation  and
condensed matter physics. An important class of systems in condensed matter physics  which one could try and study using this correspondence consists of  fermions at finite density with strong correlations.
Landau Fermi liquid theory is one paradigm that often describes such systems, but it can fail.
The resulting Non-Fermi liquid behaviour is poorly understood and believed to be of considerable interest, e.g.,
 in the study of High $T_c$ superconductors in $2+1$ dimensions.

On the gravity side,  the Einstein Maxwell Dilaton (EMD) system consisting of gravity and  a Maxwell  gauge field  coupled to a neutral scalar
(the Dilaton) is  of considerable interest from the point of view of studying this problem.
Fermions in  the boundary theory  carry a  conserved charge - fermion number- so it is natural to include a gauge field in
the bulk. The presence of a neutral scalar allows for promising new  phases  to arise
where the entropy vanishes at non-zero chemical potential and zero temperature,
as was discussed in \cite{Gubser:2009qt}, \cite{GKPT}, \cite{Kiritsis}.  These phases correspond to compressible states of matter
with unbroken fermion  number symmetry \footnote{The significance of the compressible nature of the state was emphasised to us
by S. Sachdev, see  \cite{Huijse:2011hp}.}.
It was found that the thermodynamics and transport properties of these systems, while showing the existence of gapless
excitations,  do not fit those of a Landau Fermi liquid. For example, the specific heat is typically not linear in the temperature ($T)$, at small temperatures, and the electric resistivity
also does not have the required $T^2$ dependence \footnote{
These results refer to the case when the  boundary theory is  $2+1$ dimensional with a $3+1$ dimensional
  bulk dual.}.

An exciting recent development  has shown that for an appropriate range  parameters such an EMD
  system could give rise to an
entanglement entropy which  reproduces the behaviour expected of a  system with a Fermi surface.
If we take a sufficiently big region in space in a system with a Fermi surface it is believed that the entanglement entropy goes
like
\be
\label{EEfermi}
S_{entangled} \sim A \log(A)
\ee
 where $A$ is the area of the boundary of this region \footnote{ Strictly speaking this behaviour has only been proven for
 free or weakly coupled fermions \cite{Wolf:2006zzb},\cite{quant-ph/0504151} but it is expected to be more generally true due to the locus of gapless excitations which arises in the presence of a Fermi surface.
 Additional evidence has also been obtained in \cite{0908.1724condmat.str-el}, \cite{1002.4635condmat.str-el},
\cite{1102.0350cond-mat}.}.
 The log enhancement is believed to be the tell-tale signature of a Fermi surface. Exactly such
 a behaviour was shown to arise   for appropriate choices of parameters in the EMD system
 in \cite{Ogawa}, see also \cite{Huijse}. In addition, it was argued that  the  specific heat, at small temperatures,
could be understood on the basis of    gapless excitations which  dispersed with
a non-trivial dynamical exponent.

Taken together, these developments suggest that for an  appropriate range of parameters the  EMD system
 could perhaps describe phases where a Fermi surface does form but where the resulting description is not of Landau Fermi
liquid type.  While this is a promising possibility it is far from being definitely established.
In fact, as has been known for some time now, at large $N$ (classical gravity) the system does not exhibit some
of the standard  characteristics expected of a system with a Fermi surface. For example there are no  oscillations
in the magnetisation and other properties as the magnetic field is varied  (the de Haas-van Alphen effect) ,
 nor are there any $2 k_F$
Friedel oscillation \footnote{At one loop de Hass- van Alphen type oscillations are seen, \cite{Denef:2009yy}.
For some recent discussion of Friedel oscillations in ($1+1$) dim. see \cite{Faulkner:2012gt}.}.
 More recently the non-zero momentum
current-current two point function has been calculated and found to have suppressed weight at small frequency
\cite{Hartnoll:2012wm}.

In this paper we will continue to study this class of systems from the gravity side by turning on an additional
  magnetic field and determining the resulting response. In our work the magnetic field will be kept small compared to
charge density in the boundary theory. We will be more specific about what this means in terms of the energy scales of
the boundary theory shortly. For now let us note that without a magnetic field the purely electric theory has a scaling-type
solution (more correctly a hyperscaling violating solution).
The magnetic field is kept  small so that its effects are  a small perturbation compared to the electric field in the ultraviolet
(UV) of this scaling solution.

\subsection{Key Results}


We find that   in the dilaton system  even a small  magnetic field can have an important 
effect  at long distances since the magnetic field  can become relevant  in the Infra-red (IR).  
The resulting thermodynamic and entanglement
entropy can  then change significantly. In particular this happens for the whole  range of parameters
where the entanglement entropy is of the form eq.(\ref{EEfermi}).

More specifically, the EMD system we analyse is characterised by two parameters $(\alpha,\delta)$ which are defined
in
\footnote{The relation of $(\alpha,\delta)$ to the parameters $(\theta, z)$ now more conventionally used
 in the literature is given in eq.(\ref{defztheta}). In particular
$\alpha=-3\delta$ corresponds to $\theta=d-1=1$.}
 eq.(\ref{deff}), eq.(\ref{defV}).
When $|\alpha|>|\delta|$ we show that the magnetic field is relevant in the IR
and the geometry in the deep infra-red (small values of the radial coordinate $r$ we use ) flows
to an $AdS_2\times R^2$ attractor.
As a result the system acquires a non-zero extensive entropy even at zero temperature.
The entanglement entropy also changes and grows like the volume of  the region of interest \footnote{
A potential confusion with our terminology arises because we are in two dimensions. Thus the volume of the region
of interest  is actually its area and the area of the  boundary of this region is the  perimeter.} (for large enough volume).
In particular, this happens for the values   $\alpha =-3 \delta$
 where the purely electric theory gives rise to an entanglement
of the form eq.(\ref{EEfermi}).

%

We also analyse the thermodynamics and some transport properties of the resulting state.
The system continues to be compressible in the presence of a magnetic field and its specific heat is linear at small temperatures. Both these facts indicate the presence of gapless excitations.
 In general the system has a magnetisation which is linear in the magnetic field
and which is expected to be diamagnetic.
The  $AdS_2\times R^2$ attractor leads to the magnetisation having
 a temperature dependence, at small $T$, which can become important even for small magnetic fields, eq.(\ref{esttemp}).

The summary is  that  for parameters where the
electric theory has an entanglement of the form eq.(\ref{EEfermi}),
suggesting that it is  a  non-Landau Fermi liquid, the magnetic field is a relevant perturbation in the IR.
As a result even a small magnetic field has a significant effect
on the state of the system at long distances. The state continues to be compressible, with a linear specific heat, but the thermodynamic
entropy at zero temperature is now extensive and  the entanglement entropy scales like
the volume of the region of interest, this behaviour also being linked to the extensive ground state entropy   
\footnote{More generally from the fact that
 the magnetic field is a relevant perturbation in the IR
we learn that the compressible state described by the purely electric  solution ``anti-screens''
 the effects of the magnetic field making them grow at larger distances.}.
 At intermediate length scales for which the relevant region of 
 the geometry is still reasonably well approximated by the 
hyperscaling violating type metric  and the effects of the magnetic field are small, the behavior of the 
system continues to be essentially  what it was in the absence of the 
magnetic field. In particular the thermodynamics is essentially unaffected by the magnetic field and the entanglement entropy also stays unchanged. 
Similar  results for the existence of an  $AdS_2\times R^2$ attractor and associated changes in thermodynamic and
entanglement entropy etc  are true in the  whole region where $|\alpha|>|\delta|$.

The  behaviour mentioned above  is roughly analogous to what happens  in a weakly coupled system with a 
Fermi surface \footnote{We thank the referee for his/her comments which have lead to this paragraph being incorporated in the revised version of the paper.}.
While in this case  the introduction of a small magnetic field leads to  the formation of 
 Landau levels, at intermediate energies still low compared to the Fermi energy but
big compared to  the spacing of the Landau levels, and 
correspondingly at intermediate length scales smaller than the magnetic length,  
the behaviour continues to be 
essentially that of a system with a Fermi surface. In particular the thermodynamics is essentially 
unchanged by the small magnetic field and the entanglement entropy is also expected  to have the 
$A \log(A)$ behaviour at these length scales. Going to much lower energies of order the spacing between the  Landau level
and  correspondingly to   distance scales of order or  longer  the magnetic length  
though the behaviour of the system can change. For example in the free fermion theory,  depending on the 
fermion  density,  a   Landau level  can be fully or partially filled, and  partial filling  would result
in an extensive ground state entropy. 

In other regions of parameter space where $|\alpha|<|\delta|$ the magnetic perturbation is either not relevant in the
IR and thus essentially leaves the low-energy and large distance behaviour of the system unchanged.
Or it is relevant but we have not been able   to  completely establish the resulting geometry to which
the system flows   in the deep IR.

The  paper is planned as follows. We start with a brief description of the dilatonic system and the hyperscaling violating metrics in \S2. The effects of a magnetic field are discussed in \S3. The resulting thermodynamics is discussed in \S4 and the entanglement entropy in \S5. We end with a discussion of results and some concluding comments in \S6.
Appendix A contains important details about the numerical analysis.

Before ending the introduction let us also comment on some related literature.
For a discussion of probe fermions in the extremal RN geometry and the resulting non-Fermi liquid behaviour see,
\cite{Lee:2008xf},\cite{Liu:2009dm}, \cite{Cubrovic:2009ye},\cite{Faulkner:2009wj},\cite{Hartnoll:2009ns},\cite{Faulkner:2010tq}.
The EMD system has been studied in \cite{Gubser:2009qt},\cite{GKPT},\cite{Kiritsis},\cite{Perlmutter},\cite{GIKPTW},\cite{Meyer:2011xn},\cite{Gouteraux:2011xr},\cite{IKNT},\cite{Gouteraux:2011ce},\cite{Gouteraux:2011qh}.
The subject of entanglement entropy has   received considerable attention in the
literature lately, for
a partial list of references see \cite{Bombelli:1986rw},\cite{Srednicki:1993im},\cite{Holzhey:1994we},
\cite{Calabrese:2004eu},\cite{Calabrese:2005zw},
for early work; and for a  discussion   within   the context of the AdS/CFT
correspondence
see  \cite{RT1},\cite{Ryu} .

Two papers in particular have overlap with the work reported here. While their motivations  were different the
analysis carried out in these papers is similar to ours.
The EMD system with the inclusion of possible  higher order corrections was analysed  in \cite{Kachru22}
and it was found that sometimes these  corrections  could change the behaviour of the  geometry resulting  in
an $AdS_2\times R^2$ region in the deep IR.
This analysis was generalised to the case with hyperscaling violation in
the more recent paper \cite{Jyotirmoy} which appeared while our  work   was being completed.


\section{The Dilaton Gravity System}
We work in $3+1$ dimensions in the gravity theory  with an action
\be
\label{action1}
S=\int d^4x\sqrt{-g}[R-2(\nabla \phi)^2  -f(\phi) F_{\mu\nu}F^{\mu\nu} -V(\phi)].
\ee

Much of our emphasis will be on understanding the near horizon region  of the black brane solutions which arise in this system. This region is more universal, often having the properties of an attractor,
and also determines the IR behaviour of the system.
In this region, in the solutions of interest, the dilaton will become large, $\phi \rightarrow \pm \infty$.
The potential and gauge coupling function take the approximate forms
\begin{eqnarray}
f(\phi) & = & e^{2\alpha \phi} \label{deff} \\
V&= & -|V_0| e^{2\delta \phi} \label{defV}
\end{eqnarray}
 along this direction of field space.
 The two parameters,    $\alpha, \delta$,
  govern the behaviour of the system. For example the thermodynamic and transport properties and also
the entanglement properties crucially depend on these parameters.
The action in eq. (\ref{action1}) has a symmetry under which the sign of $\phi, \alpha, \delta$ are reversed.
Without loss of generality we  will therefore  choose $\delta>0$ in the discussion which follows.

Our analysis will build on the earlier investigations in \cite{Kiritsis} and \cite{IKNT},
and our conventions will be those in \cite{IKNT}.
We will work in coordinates where the metric is,
\be
\label{ansatz1}
ds^2 = - a(r)^2 dt^2 + { dr^2 \over a(r)^2 } + b(r)^2 (dx^2 + dy^2)
\ee
The horizon of the extremal  black brane will be taken to lie at $r=0$.

The
gauge field equation of motion gives,
\be
\label{gf}
 F = {Q_e \over f(\phi) b^2 }  dt \wedge dr + Q_m dx \wedge dy.
\ee
The remaining equations of motion can be conveniently expressed in terms of an
effective potential  \cite{GIJT}
\begin{equation}\label{Veff}
V_{\text{eff}} = {1 \over b^2} \left( e^{-2 \alpha \phi} Q_e^2  + e^{2 \alpha
\phi} Q_m^2 \right) - {b^2 |V_0|  \over 2} e^{2 \delta \phi},
\end{equation}
and are given by,
\begin{align}
 \label{em1}
 (a^2 b^2)'' = & 2 |V_0| e^{2 \delta \phi} b^2 \\
 \label{em2}
 {b'' \over b} = & -  \phi'^2 \\
 \label{em3}
 (a^2 b^2  \phi')'   = &   {1 \over 2} \partial_\phi V_{\text{eff}} \\
\label{em4}
 a^2 b'^2 + {1 \over 2} {a^{2}}' {b^{2}}' = &  a^2 b^2 \phi'^2 - V_{\text{eff}}.
\end{align}

\subsection{Solutions With Only Electric Charge}
\label{pureelectric}
Next let us briefly review the solutions with $Q_m$ set to zero which carry only electric charge.
The solution in the near-horizon region take the form,
\begin{equation}\label{ansatz2}
a = C_a r^{\gamma} \hspace{10mm} b= r^\beta \hspace{10mm} \phi = k \log{r}
\end{equation}
where the coefficients $C_a, \gamma, \beta, k$ and the electric charge $Q_e$ are given by
\begin{align}\label{case1}
\beta = { (\alpha+\delta)^2 \over 4 + (\alpha+\delta)^2}   \hspace{10mm}
\gamma = 1 -{ 2 \delta (\alpha+\delta) \over 4 + (\alpha+\delta)^2}
&\hspace{10mm}
 k = - { 2 (\alpha+\delta) \over 4 + (\alpha+\delta)^2}
\\
\label{case11} C_a^2  = |V_0| {\left( 4 + (\alpha+\delta)^2 \right)^2 \over 2
\left(2 + \alpha (\alpha+\delta) \right) \left( 4 + (3 \alpha-\delta)
(\alpha+\delta) \right)} & \hspace{5mm} Q_e^2 = |V_0|{ 2 - \delta (\alpha+\delta)
\over 2 \left( 2 + \alpha (\alpha+\delta) \right)}.
\end{align}

It might seem strange at first that  the electric charge $Q_e$ is fixed, this happens
 because in the  near-horizon metric we work with the
the time (and spatial coordinates)  which have been rescaled
compared to their values  in the UV.

\begin{figure}[h]
\begin{center}
\includegraphics[scale=.8]{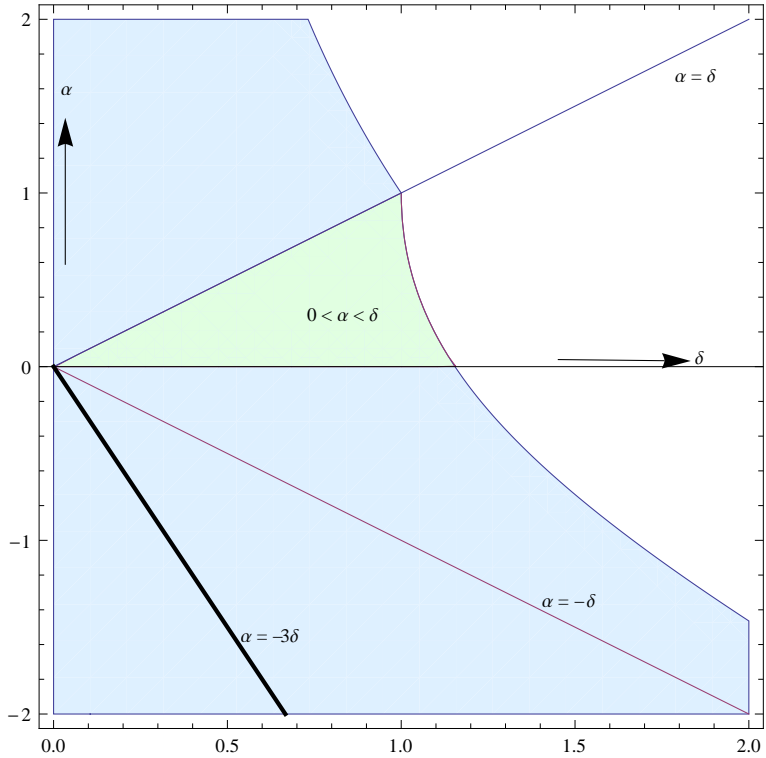}
\caption{The blue and green shaded regions are  allowed by the 
various positivity and thermodynamics constraints
for the electric scaling solutions. The straight lines in the $(\delta,\alpha)$ plane  
 demarcate various regions which  will be relevant for  the discussion in the following sections.}~\label{RegionPlot1}
\end{center}
\end{figure}

The following three  conditions must be satisfied for this solution to be valid
: $Q_e^2>0, C_a^2 >0, \gamma>0$. These give the constraints,
\begin{eqnarray}\label{constraint1}
2 - \delta (\alpha+\delta)&>& 0 \\
\label{constraint2}
2 + \alpha (\alpha+\delta) &>& 0 \\
\label{constraint3} 4 + (3 \alpha-\delta) (\alpha+\delta) &>& 0 \\
\label{constraint4}
 4 + (\alpha - 3 \delta) (\alpha+ \delta)&>&0.
\end{eqnarray}
The last of these  conditions follow from the requirement that
\be
\label{condgamma}
2\gamma-1 >0
\ee
so that  the specific heat is positive.
Figure (\ref{RegionPlot1}) shows the the region in the $(\delta,\alpha)$ plane, with $\delta>0$,
allowed by the above constraints.

To summarise our discussion,  the metric in the purely electric solution takes the form
\be
\label{finalmet}
ds^2=-C_a^2 r^{2\gamma}dt^2 + {dr^2 \over  C_a^2 r^{2\gamma}} + r^{2\beta} (dx^2+dy^2).
\ee
And the dilaton is given in eq.(\ref{ansatz2},\ref{case1})
While this solution  is not scale invariant  it does admit a conformal killing vector.
This follows from noting that under the transformation
\begin{eqnarray}
r & = & \lambda \tilde r  \label{rscale}\\
t & = & \lambda^{1-2 \gamma} \tilde t \label{tscale}\\
\{x,y\} &=& \lambda^{1-\gamma-\beta} \{\tilde x,\tilde y\} \label{xyscale}
\end{eqnarray}
the  metric eq.(\ref{finalmet}) remains invariant upto a overall scaling,
\begin{equation}
ds^{2} = \lambda^{2-2\gamma} \{ -C_a^2 \tilde r^{2 \gamma} d \tilde
t^{2}+\frac{d\tilde r^{2}}{C_a^2 \tilde r^{2 \gamma}}
+ \tilde r^{2 \beta} (d\tilde x^{2} +d \tilde y^{2}) \}.
\end{equation}
The dilaton also  changes under this rescaling  by an additive constant,
\begin{equation}
 \phi= k ~ln(\tilde r)+ k ~ ln(\lambda)
\end{equation}
The two exponents $\gamma, \beta$ which appear in the metric are related to the dynamic exponent with which gapless
excitations disperse and hyperscaling violations, as was explained in \cite{Huijse}.

Under the coordinate change,
\begin{eqnarray}
 r &=& \tilde{r}^{-{1\over \beta}} \\
 t &=& {1 \over \beta C_a^2} \tilde{t} \\
 (x,y) &=& {1 \over \beta C_a} (\tilde{x},\tilde{y}),
\end{eqnarray}
the metric eq.(\ref{finalmet}) becomes
\begin{equation}
 ds^2 = {1 \over \beta^2 C_a^2}{1 \over \tilde{r}^2 } \left\lbrace {-{d\tilde{t}^2 \over \tilde{r}^{ 4 (z-1) \over 2-\theta} }} +
 \tilde{r}^{2 \theta \over 2 - \theta} d \tilde{r}^2 + d \tilde{x}^2 + d \tilde{y}^2 \right\rbrace.
\end{equation}
Where,
\begin{equation}
\label{defztheta}
 z= {2\gamma-1 \over \beta + \gamma -1} \hspace{20mm} \theta= {2(\gamma-1) \over \beta + \gamma -1}.
\end{equation}
This is the form of  the metric discussed  in \cite{Huijse} (upto the  overall  ${1 \over \beta^2 C_a^2}$ factor which was set to unity by a choice of scale).
The exponent $z$ is the dynamic exponent, as we can see from the scaling weights of the $t$ and $x,y$ directions in
eq.(\ref{tscale}), eq.(\ref{xyscale}).
The exponent $\theta$ is the  hyperscaling violation exponent,  we will also explain this further in
 \S4.

Let us end this section with some more comments.
In eq.(\ref{gf})  the  two-form $F$ is dimensionless, so that $Q_e, Q_m$ have dimensions of $[Mass]^2$.
The chemical potential $\mu$   is related to $Q_e$ by
\be
\label{relqe}
Q_e \sim \mu^2
\ee
 and has dimensions of mass.

The  near-horizon geometry of the  type being discussed here can be obtained by
starting from an asymptotically AdS space in the UV for a suitable choice of the  potential $V(\phi)$.
This was shown, e.g., in  \cite{IKNT}, for additional discussion see Appendix A.
It is simplest to consider situations where the asymptotic AdS space has only one scale, $\mu$, which characterises
both the chemical potential of the boundary theory and any breaking of conformal invariance due to a non-normalisable mode
for the dilaton being turned on. In our subsequent discussion we will have such a situation in mind and the scale $\mu$ will often enter the discussion of the thermodynamics and entanglement.


Also note that the
parameter $N^2$ which will  enter for example in the entropy  eq.(\ref{entern})  is  given in terms of the
potential eq.(\ref{defV})  by
\be
\label{valn}
N^2\sim {1\over G_N |V_0|}.
\ee
Again to keep the discussion simple we will take the cosmological constant for the asymptotic $AdS$ to be of order $V_0$
so that $N^2$ is also number  of the degrees of freedom in the UV\footnote{
The  examples studied in \cite{IKNT} are of this type. There  the full potential was taken to be
$V(\phi)=-2 |V_0| \cosh(2\delta \phi)$ (see eq.(F.1 of \cite{IKNT}). As a result in the asymptotic region $r\rightarrow \infty$ the potential goes to its maximum value,
 $V \rightarrow V_{\infty} = -2 |V_0| \sim -|V_0|$.}.


The solutions we have considered can have curvature singularities as $r\rightarrow 0$,  when such singularities are absent
tidal forces can still diverge near the horizon, e.g., see  \footnote{Sometimes the geometries can be regular
with no singularities or diverging tidal forces, \cite{Shago}.}, \cite{Horowitz:2011gh}.  These divergences can be cut-off by
 heating the system up to a small temperature as discussed in \cite{GKPT}, \cite{IKNT}. Also, as we will see shortly,
adding a small magnetic field can alter the behaviour of the geometry in the deep IR again removing the singular region.

\section{The Effect of the Magnetic field}
Now that we have understood the solutions obtained with only  electric charge  we are ready to study the effects of adding a small
magnetic field.

The presence of the magnetic field gives rise to an additional term in the  effective potential eq.(\ref{Veff}).
The  magnetic field is a small perturbation if this term is small compared to the electric charge term, giving rise to the condition
\be
\label{condsmag}
\left({Q_m\over Q_e}\right)^2 \ll e^{-4\alpha \phi}.
\ee
From eq.(\ref{ansatz2}) and eq.(\ref{case1}) we see that
\be
\label{condcs1a}
e^{-4 \alpha \phi}=r^{-4 \alpha k}
\ee
so   eq.(\ref{condsmag}) in fact  gives rise to a condition on the radial coordinate
\be
\label{condrc}
\left({Q_m\over Q_e}\right)^2 \ll r^{-4 \alpha k}.
\ee

By a small magnetic field we mean more precisely choosing a suitable value of $Q_m$ and starting at  a value  of
the  radial coordinate $r$ where eq.(\ref{condrc}) is met.
We will be then interested in asking if this magnetic perturbation continues to be small in the IR, i.e., even smaller values
of $r$, or if its effects grow \footnote{As is clear from eq.(\ref{condrc}) and we will study this shortly in more detail,
the magnetic field is relevant in the IR when $\alpha k<0$. It is easy to see from eq.(\ref{case1}), eq.(\ref{case11})
that when this condition is met the coupling $g^2=e^{-2\alpha \phi}$
 in the  purely electric solution is weakly
coupled in the IR since  $g^2 \rightarrow 0$ as $ r\rightarrow 0$.}.

The requirement that the magnetic field is small can be stated more physically as follows.
Consider a purely electric solution which asymptotes to $AdS$ space in the far UV and let $\mu$ be the only
scale characterising the boundary theory which is dual to this  electric theory as
 discussed in the previous section.
Then the magnetic field is small if it satisfies the condition
\be
\label{condmag}
|Q_m|\ll \mu^2,
\ee
 so that its effects can be neglected in  the UV and continue to be small all the way to the
electric scaling region.

Our discussion  breaks up into different cases depending on the values of the parameters $\alpha,\delta$.
We will choose $\delta \ge 0$ in the discussion below without any loss of generality.
Let us also note that although we do not  always mention them for an electric solution  to exist the additional conditions
eq.(\ref{constraint1})-eq.(\ref{constraint4})
must also be met.

We now turn to the various cases.

\subsection{ Case I. $-\delta < \alpha<0 $}

In this case the magnetic perturbation is irrelevant in the infrared.
From eq.(\ref{case1})
 we see that $ \alpha k  > 0   $ so that
\be
\label{condcs1b}
e^{-4 \alpha \phi}=r^{-4 \alpha k} \rightarrow \infty
\ee
as $r\rightarrow 0$.
Thus  choosing a value of $Q_m, r,$ where  eq.(\ref{condsmag}) is met and going to smaller values of $r$,
 eq.(\ref{condsmag}) will
continue to hold and therefore the effects of the magnetic field will continue to be small \footnote{On the other hand the
magnetic field gets increasingly more important at large $r$, i.e., in the UV. However from numerical solutions one sees that
 for a suitable $V(\phi)$, when $Q_m^/\mu^2 \ll 1$ its effects   continue to be small all the way upto the asymptotic AdS region.}.
In this range of parameters then the low temperature behaviour of the system and its low frequency
response will be unchanged from the purely electric case.
Also the entanglement entropy in the boundary theory of a region of  sufficient large volume  will be
 unchanged  and be  given as we shall see in \S5 by   eq.(\ref{entab}).

\subsection{Case II. $|\alpha| > \delta $}
\label{subsectioncase2}
In this case the magnetic perturbation is relevant in the infrared and in the deep infrared the solution approaches
an attractor of the extremal RN type. The dilaton is drawn to a fixed and finite value $\phi_0$ and
does not run-away and  the near-horizon geometry is  $AdS_2\times R^2$ with    the metric components
eq.(\ref{ansatz1}), being
\begin{eqnarray}
b & = & b_0  \label{defb0}\\
a^2 & = & {r^2\over R_2^2} \label{defC},
\end{eqnarray}
where $b_0, R_2^2$ are constants with $R_2$ being the radius of $AdS_2$..
Note that in this attractor region of the spacetime  the effects of the electric and magnetic
fields are comparable.

To establish this result  we   first show  that  eq.(\ref{em1}), eq.(\ref{em2}), eq.(\ref{em3}) and eq.(\ref{em4})
 allow for such an  attractor solution.
Next, starting with this  attractor solution we identify appropriate perturbations and establish numerically that the
solution flows to the electric scaling solution in the UV.

It is easy to check that the equations of motion allow for a
 solution of the type described above.  Eq.(\ref{em3}) and eq.(\ref{em4})
 are met with $\phi$ being constant and $b$ being constant as long as
the conditions  $V_{eff}=\partial_\phi V_{eff}=0$ are met.
This gives rise to the  conditions
\begin{eqnarray}
 e^{-2 \alpha \phi_0} Q_e^2 + e^{2 \alpha \phi_0} Q_m^2 &=& {b_0^4  |V_0| \over
2} e^{2 \delta \phi_0} \label{atone} \\
 -  e^{-2 \alpha \phi_0} Q_e^2 + e^{2 \alpha \phi_0} Q_m^2  &=& \left( {\delta
\over \alpha} \right) {  b_0^4  |V_0| \over 2} e^{2 \delta \phi_0} \label{attwo}
\end{eqnarray}
which determine $\phi_0, b_0$.
Eliminating $b_0$ between the two equations gives
\begin{equation}
\label{atphi}
 e^{4 \alpha \phi_0} = { Q_e^2 \over Q_m^2} \ {1 + {\delta \over \alpha} \over 1
- {\delta \over \alpha} }
\end{equation}
The  LHS must be  positive, this gives a constraint $|{\delta \over \alpha}| <1$ which is indeed
true  for Case II.
Substituting eq.(\ref{atphi}) in eq.(\ref{atone})  next
determines $b_0$ in terms of   $\phi_0$ to be
\begin{equation}
\label{valb0}
 b_0^4  =  {4 Q_e^2 e^{-2 \phi _0 (\alpha + \delta) } \over |V_0| \left( 1 -
{\delta \over \alpha}  \right) }.
\end{equation}
Of the remaining equations eq.(\ref{em2}) is trivially satisfied while eq.(\ref{em1}) determines $R_2$ to be
\be
\label{valr2a}
R_2^2={1\over |V_0|}(|{\alpha-\delta\over \alpha+\delta}|)^{\delta\over 2 \alpha}({Q_m^2\over Q_e^2})^{\delta \over 2 \alpha}.
\ee
We see that  for $\alpha>0$, $R_2 \rightarrow 0$  as $Q_m\rightarrow 0$, making the $AdS_2$ highly curved,
while for $\alpha<0$, $R_2\rightarrow \infty$ as $Q_m\rightarrow 0$.

Appendix A contains some discussion of the two  perturbations in this $AdS_2\times R^2$ solution which grow in the UV.
Starting with an appropriate choice of these two perturbations we find that the solution flows to the
electric scaling solution in the UV. This can be seen in Fig. (\ref{abvsr}) and (\ref{phivsr1}).
For an appropriate choice of potential going out even further in the UV one finds that  the solution becomes asymptotically
$AdS_4$,
as shown in Fig. (\ref{logabvsr}) and (\ref{phibvsr2}).

The $AdS_2\times R^2$ near-horizon geometry changes the IR behaviour of the system completely.
As discussed in the introduction there is now an extensive thermodynamic entropy and the
entanglement entropy also scales like the volume, for large enough volume.
For additional discussion of the thermodynamics see \S4 .

\subsection{Case III. $0<\alpha<\delta$}
\label{case3}
In this case also  we will see that
 the magnetic perturbation is relevant in the IR.  Our analysis for what the end point
is in the IR will not be complete, however.

We do
 identify
 a candidate ``run-away'' attractor as the IR end point of the system.
In this attractor solution the magnetic field dominates and the effects of the electric field
are negligible in comparison. As a result  a  solution
 taking the hyperscaling violating form eq.(\ref{ansatz2}), for an  appropriate choice of exponents, exists.
We will refer to this solution as the magnetic scaling solution below.
Unfortunately, we have not  been able to  satisfactorily establish
 that starting with the electric solution of interest one
does indeed end in this magnetic scaling solution in the IR.
This requires additional numerical work.

To see that the magnetic perturbation is relevant in the IR note that $\alpha k<0$ in this region
so that eventually, for small enough values of $r$, condition eq.(\ref{condrc})  will no longer hold and the effects of the magnetic field will become significant.

 To identify the candidate run-away attractor let us begin  by noting that the effective potential eq.(\ref{Veff}) and thus the
equations of motion are invariant under the transformation,
$Q_m \leftrightarrow Q_e$ accompanied by  $\alpha \rightarrow -\alpha$ with the other parameters staying the same.
Under this transformation the region discussed in Case I maps to the region $0<\alpha<\delta$.
The discussion for Case I above  then shows that,  with $Q_m$ present, in this region of parameter space there is a
consistent solution where the effects of the electric charge in the deep IR can be neglected.
The solution takes  the form, eq.(\ref{ansatz2}) and eq.(\ref{case1}), eq.(\ref{case11}), with $Q_e\rightarrow Q_m$ and
$\alpha\rightarrow -\alpha$. This is the magnetic scaling solution referred to above.
Actually,
this solution exists only if  $(-\alpha,\delta)$ meet the conditions
 eq.(\ref{constraint1})-eq.(\ref{constraint4}). In  Fig.(\ref{RegionPlot1}) the region for Case III where all the conditions eq.(\ref{constraint1})-eq.(\ref{constraint4}) are met is shown in green. 
It is easy to  see that for any point in this allowed  green region the corresponding point
 $(-\alpha,\delta)$ automatically lies in the allowed blue region.

Assuming that we have identified the correct IR end point
  we see that the thermodynamic entropy at extremality continues to
vanish once the magnetic perturbation is added. It is also easy to see that the entanglement entropy is of the
form eq.(\ref{entab}).

A more complete analysis of the  system in this region of $(\alpha,\delta)$ parameter space is left for the future.

\subsection{Additional Comments}
\label{addcom}
We end this section with some  comments.
It is sometimes useful to think of the  solutions we have been discussing as being embedded in a more complete one
which asymptotes to AdS space in the UV. The dual field theory then lives on the boundary of AdS space and standard rules of
AdS/CFT can be used to understand its behavior. We take this theory to have one scale, $\mu$, as discussed in \S2.
  In addition the magnetic field is also turned on with $Q_m/\mu^2 \ll 1$.
The full metric for this solution will be of the form eq.(\ref{ansatz1})  and in the UV will become
AdS space \footnote{We have set $R_{AdS}=1$.} :
\be
\label{metadsa}
ds^2=-r^2 dt^2 + {1\over r^2} dr^2 + r^2(dx^2+dy^2)
\ee
Starting with this geometry for $r\rightarrow \infty$ it will approach the electric scaling
solution when $r \lesssim  \mu$.

The magnetic field becomes a significant effect when its contribution to the effective potential eq.(\ref{Veff})
 is roughly comparable to the electric field.  This gives a condition  for the dilaton
\be
\label{condcom}
e^{-4 \alpha \phi}\sim {Q_m^2\over Q_e^2}
\ee
Using eq.(\ref{case1})   this happens at a radial location $r\sim r_*$ where
\be
\label{valrs}
r_*\sim  \mu \left({Q_m^2 \over Q_e^2}\right)^{-1\over 4 \alpha k}
\ee
Here we have introduced the parameter $\mu$ which was set equal to unity in eq.(\ref{ansatz2}),
eq.(\ref{case1}), eq.(\ref{case11}).
For Case II and III where the magnetic perturbation is relevant in the IR, $\alpha k <0$, and the magnetic field continues
to be important for all $r<r_*$. In Case II for $r\ll r_*$ the solution becomes $AdS_2\times R^2$.
In Case III we have not identified the IR endpoint with certainty when $r\ll r_*$.
For Case I  the magnetic perturbation is irrelevant in the IR.

\begin{figure}[h]
\begin{center}
\includegraphics[scale=.6]{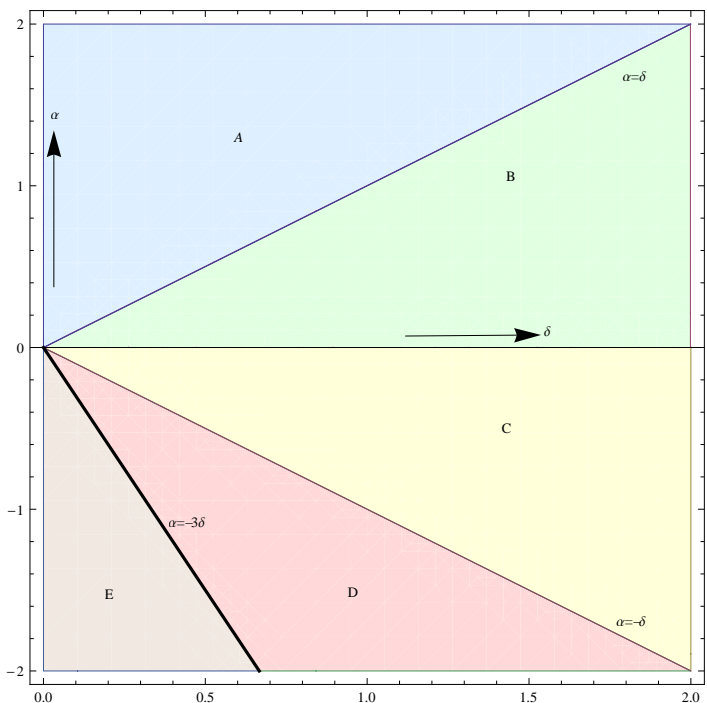}
\caption{Figure showing various region in $(\delta,\alpha)$ space. Details of these regions can be found in Table (1)}
~\label{RegionPlot2}
\end{center}
\end{figure}

Second, Figure (\ref{RegionPlot2}) shows a plot of various regions in the  $(\delta,\alpha)$ plane, with $\delta>0$.
Region C corresponds to Case I. Regions A, D and E correspond to Case II. And region B corresponds to Case III.
The line $\alpha=-3\delta$ which is of special interest is the thick black line separating regions E and D.
These regions are also described in terms of the parameters $\beta,\gamma$, eq.(\ref{case1}), eq.(\ref{case11})
in Table (\ref{table1}). The corresponding values of the parameters $(\theta,z)$ can be obtained from
eq.(\ref{defztheta}).
\begin{table}[h]
 \centering
\begin{tabular}{|c|c|c|}
 \hline
Region &  $\delta$,$\alpha$ & $\beta$, $\gamma$ \\
\hline
A & $\alpha>\delta$ & $1+\beta>\gamma>1-\beta$\\
\hline
B & $0<\alpha<\delta$& $\gamma<1-\beta$\\
\hline
C & $0>\alpha>-\delta$&  $\gamma<1-\beta$\\
\hline
D & $-3 \delta<\alpha<-\delta$& $\gamma>1+\beta$\\
\hline
E & $\alpha<-3 \delta$ & $1+\beta>\gamma>1-\beta$\\
\hline
\end{tabular}
\caption{Various Regions in ($\delta,\alpha$) space}\label{table1}
\end{table}
\section{More on Thermodynamics}
In this section we will discuss the thermodynamic behaviour in the presence of the magnetic field in some more detail.
%
The introduction of a small magnetic field in the dilaton  system  can have
a significant  effect on the IR behaviour as we have already discussed.
 Here we will study some  additional aspects of the  resulting thermodynamics.

In our system the role of the Fermi energy is played by the chemical potential $\mu$.
Let us start with the purely electric theory first at a small temperature $T$
\be
\label{stemp}
T/\mu \ll 1
\ee
As discussed in \cite{Kiritsis}, \cite{IKNT} the entropy density $s=S/V$  goes like
\be
\label{enta}
s\sim N^2 T^{2\beta\over 2\gamma-1}.
\ee
Under the scaling symmetry equations, (\ref{rscale}), (\ref{tscale}), (\ref{xyscale}),
 $s$ has dimensions of $L^{\theta -2}$, where $\theta$ is defined
in eq.(\ref{defztheta}) and
$L$ transforms in the same way as  the $(x,y)$ coordinates do in eq.(\ref{xyscale}).
Thus  $\theta$ is the exponent related to
hyperscaling violation.

Now we can consider introducing a small magnetic field $Q_m$. Since the stress energy of the electromagnetic field
is quadratic in $Q_m^2$ this should result  in a correction to the entropy which is of order $Q_m^2$.
The scaling symmetry eq.(\ref{tscale}), eq.(\ref{xyscale})  then fixes the resulting
temperature dependence of this correction so that  $s$ is given by
\be
\label{corrs}
s\sim N^2 \mu^2 \left({T \over \mu}\right)^{2\beta\over 2\gamma-1}\left(1+ s_1 \left({Q_m \over \mu^2}\right)^2
\left({T\over \mu}\right)^{4 \alpha k \over 2 \gamma-1}\right)
\ee
where $k$ is defined in eq.(\ref{ansatz2}), eq.(\ref{case1}) and $s_1$ is a $\mu$ independent constant.
We see that the magnetic field can be regarded as a small perturbation only for temperatures meeting the condition
\be
\label{condsmallqm}
\left({Q_m \over \mu^2}\right)^2 \left({T\over \mu}\right)^{4\alpha k \over 2\gamma-1}\ll 1
\ee
We have numerically verified that the coefficient $s_1$ indeed does not vanish  for generic values of
$(\alpha,\delta)$.

The condition eq.(\ref{condsmallqm})  is in agreement with the discussion of section \S2 where we
found that the
 magnetic field is irrelevant or relevant in the IR depending on the sign of  $\alpha k$.
 Since $2\gamma-1 > 0$, eq.(\ref{condgamma}), we see from
eq.(\ref{corrs}) that when $\alpha k > 0$ the effects of the magnetic field on the
entropy vanish as $T\rightarrow 0$. On the other hand when $\alpha k<0$ these effects grow as
 $T\rightarrow 0$.

\subsection{More on  Case II.}
One region of the parameter space where $\alpha k <0$ corresponds to Case II. As  discussed in \S2
in this case the resulting  geometry  for $T=0$ in the deep IR is
of the extreme RN type  and the entropy at extremality does not vanish.
 From eq.(\ref{valb0}) this entropy is given by
\be
\label{entern}
S=s_0 V N^2 \mu^2 \left({Q_m \over \mu^2} \right)^{\alpha+\delta \over 2 \alpha}
\ee
where $s_0$ is a dimensionless constant, $V$ is the volume and we have used eq.(\ref{relqe}).
The remaining region of parameter space where $\alpha k <0$ corresponds to Case III.
For this case as discussed  in \S2  our analysis is not complete. If the  IR in the gravity theory
 is an attractor of the magnetic scaling type described in \ref{case3}
 then the entropy vanishes at extremality.

It is also worth commenting on the behaviour of some of the other thermodynamic variables for Case II.
We start with the case where both $Q_m,T$ vanish, then first introduce a small $Q_m/\mu^2 \ll 1$ and finally a small
temperature. The temperature we consider meets the condition $T/\mu \ll 1$. In fact it is taken small enough
to meet the more stringent condition
\be
\label{mst}
{Q_m^2\over \mu^4}\gg \left({T \over \mu}\right)^{-4\alpha k \over 2 \gamma-1}
\ee
so that eq.(\ref{condsmallqm}) does not hold and the near horizon geometry is that of a near- extremal RN black brane
at a small non-zero temperature.

 The discussion of thermodynamics  is conceptually simplest if we think of the gravity solution being asymptotic in the
deep UV to  AdS space with a possible non-normalisable mode for the dilaton turned on, as was discussed in \ref{pureelectric}.
In the absence of a magnetic field the dual field theory is a relativistic theory with the  coupling constant dual to the
dilaton being turned on and thus scale invariance being broken. The energy density $\rho$ and pressure $P$ for such a system
at zero temperature are
given by
\begin{eqnarray}
\rho & = & c_1 N^2 \mu^3 + \rho_0 \label{rhothermo} \\
P & = & {c_1 \over 2} N^2 \mu^3 - \rho_0 \label{pthermo}
\end{eqnarray}
where the $\rho_0$ term arises due to the cosmological constant induced by to the breaking of scale invariance when the
 non-normalisable mode of the dilaton is turned on.

On introducing a small magnetic field the  geometry changes for Case II significantly in the deep IR.
However one expects that the resulting change in $\rho ,p$,  which are determined by the normalisable mode of gravity at the boundary, is small.
 Since the stress-energy in the
bulk changes at quadratic order in $Q_m$, as was discussed above, this correction should be of order $Q_m^2$.
Thus the pressure, working still at zero temperature,   would become
\be
\label{presst}
P= {c_1 \over 2} N^2  \mu^3 - \rho_0+ a_1 N^2 {Q_m^2 \over \mu}
\ee
where $a_1$ is a dimensionless constant.
The resulting magnetisation can be obtained using the thermodynamic relation
\be
\label{thermorel}
SdT+Nd\mu-VdP+ MdQ_m=0.
\ee
Keeping $T=0$ and $\mu$ fixed gives
\be
\label{maga}
{M \over V}= {dP \over dQ_m} = 2 a_1 N^2 {Q_m\over \mu}
\ee
We expect this magnetisation to be diamagnetic.

Introducing a small temperature next will result in a temperature dependence in the pressure and the magnetisation.
The change in the pressure keeping $\mu, Q_m$ fixed and increasing $T$ slightly is given from eq.(\ref{thermorel}) by
\be
\label{chngp}
\Delta P=\int s dT  = s_0 N^2 \mu^2 \left({Q_m \over \mu^2}\right)^{\alpha+\delta \over 2 \alpha} T
\ee
where we have used eq.(\ref{entern}).
Adding this to eq.(\ref{presst}) gives the total pressure to be
\be
\label{presstb}
P= {c_1 \over 2}  N^2 \mu^3 - \rho_0+ a_1 N^2 {Q_m^2 \over \mu} + s_0 N^2 \mu^2 \left({Q_m \over \mu^2}\right)^{\alpha+\delta \over 2 \alpha} T
\ee
The resulting magnetisation also acquires a linear dependence on temperature
\be
\label{magb}
{M\over V}=2 a_1 N^2 {Q_m\over \mu} + s_o N^2 \left({\alpha+\delta \over 2 \alpha}\right) T
\left({Q_m\over \mu^2}\right)^{\delta-\alpha\over 2 \alpha}
\ee
Notice that in Case II $|\alpha| > \delta$ and therefore the exponent ${\delta-\alpha\over 2 \alpha}$ in the second term
on the RHS is negative. Since $Q_m/\mu^2 \ll 1$ this means that the coefficient of the term linear in $T$ in the
magnetisation is enhanced. As a result at a  small temperature of order
\be
\label{esttemp}
{T\over \mu}\sim \left({Q_m\over \mu^2}\right)^{ 3 \alpha-\delta \over 2 \alpha}
\ee
this term will become comparable to the zero temperature contribution.

A case of  particular interest is  when  $\alpha=-3\delta$.
This corresponds to   $\theta=1$,  eq.(\ref{defztheta}), and gives rise to the logarithmic
enhancement of entropy eq.(\ref{EEfermi}). The pressure and magnetisation etc can be obtained for this case
 by substituting this relation between $\alpha,\delta$
 in the  the equations above.

Let us end this section some comments.
It is important to note that after turning on the magnetic field the state is still compressible.
The compressibility is defined by $\kappa=-{1\over V}{\partial V \over \partial P}|_{TQ_mN}$ and can be related to the change in
charge or number  density $n$ as $\mu$ is changed,
\be
\label{defnkappa}
\kappa= {1\over n^2} ({\partial n \over \partial \mu})|_{TQ_m}
\ee
From eq.(\ref{thermorel})  and eq.(\ref{presstb})  we see that
\be
\label{defnn}
n={\partial P \over \partial \mu}|_{TQ_m} = {3\over 2} c_1 N^2 \mu^2 + \cdots
\ee
where the first term on the RHS arises from the first term in $P$ in eq.(\ref{presstb}) and the ellipses denote corrections
which are small.
Thus the charge density is only slightly corrected by the addition of $Q_m$ and therefore the state remains compressible.
Our discussion above for the magnetisation etc  has been  for Case II. The analysis in case I where the magnetic field is irrelevant in the IR is straightforward. For Case III we do  not have a
 complete analysis of what happens in the gravity theory in the deep IR. A candidate  attractor
 was identified  in \S3, if this
attractor is indeed the IR end point then starting from it
the resulting thermodynamics can be worked out at small $Q_m,T$ along the lines above.

\section{Entanglement Entropy}
The entanglement entropy for the hyperscaling violating metrics we have been considering has already been worked out in
in \cite{Ogawa}, see also \cite{Huijse}. Knowing these results, the behaviour of the entanglement entropy  for our system of interest,  in the presence of a small magnetic field, can be easily deduced.

To keep the discussion self contained we first  review the  calculation of the entanglement entropy
 for  hyperscaling violating metrics
and then turn to the  system of interest.

\subsection{Entanglement Entropy in Hyperscaling Violating Metrics}
We will be considering a metric of the form
\be
\label{metent}
ds^2=-r^{2\gamma} dt^2  + {dr^2 \over r^{2\gamma}} + r^{2\beta} (dx^2+dy^2)
\ee
(this is the same form as eq.(\ref{ansatz2}) except that we have dropped the constant $C_a^2$ by appropriately scaling the metric).
In the discussion below it will be useful to think of this metric as arising in the IR starting with an AdS metric in the UV.
This could happen for an appropriately chosen potential as was discussed in \S2, \cite{IKNT}.
The field theory of interest then lives on the boundary of AdS space.
For simplicity we will restrict ourselves to a circular region ${\bf R}$  in the field theory of radius $L$.
The boundary of this region in the field theory ${\bf \partial R}$  is a circle of radius $L$.
To compute the entanglement entropy of ${\bf R}$   we work on a  fixed constant time slice and find  the
 surface in the bulk which has minimum area subject to the condition that it terminates in  ${\bf \partial R}$  at the boundary
of AdS. The entanglement entropy is then given by \cite{RT1},\cite{Ryu}.
\be
\label{EE3}
S_{EE}={A_{min}\over 4 G_N}
\ee
where $A_{min}$ is the area of this  surface.

We will work in a coordinate system of the form eq.(\ref{ansatz1})  which as $r\rightarrow 0$ becomes eq.(\ref{metent}) and
as $r\rightarrow \infty$ becomes AdS space
\be
\label{metads}
ds^2=r^2  (-dt^2) + { dr^2\over r^2} + r^2 (dx^2+dy^2).
\ee
 Replacing $(x,y)$ by $(\xi, \theta)$ the metric eq.(\ref{ansatz1})  can be rewritten  as
\be
\label{metf2}
ds^2 = -a^2  dt^2 + {dr^2 \over a^2} + b^2 (d\xi^2 + \xi^2 d\theta^2).
\ee

We expect the  minimum area
bulk surface to  maintain the circular symmetry of the boundary circle.
Such a  circularly symmetric surface has area
\be
\label{acs}
A_{bulk}= 2 \pi \int \sqrt{b^2 \left({d\xi \over dr}\right)^2 + {1\over a^2}} ~\xi(r) b(r) dr
\ee
where $\xi(r)$ is the radius of the circle which varies with $r$.
To obtain $A_{min}$ we need to minimise $A_{bulk}$ subject to the condition that as $r\rightarrow \infty$,  $\xi \rightarrow L$.
The resulting equation for $\xi(r)$ is
\be
\label{exi}
{d \over dr}\left({b^3 \xi {d \xi \over dr}\over \sqrt{b^2 ({d\xi \over dr})^2 + {1\over a^2}}}  \right) -
\sqrt{b^2 \left({d\xi \over dr}\right)^2 + {1\over a^2}} ~b =0.
\ee

Let us note that the circle on the boundary of ${\bf R}$ has area\footnote{As mentioned earlier we will persist in calling this the area although
it is of course  the perimeter.}
\be
\label{defa}
A=2\pi L
\ee

It is easy to see that as $r\rightarrow \infty$ the $b^2 ({d\xi \over dr})^2$ term in the square root in eq.(\ref{acs}) cannot
dominate over the ${1\over a^2}$  term. The contribution to $A_{bulk}$ from the $r\rightarrow \infty$ region can  then be
 estimated  easily to give
\be
\label{cba}
\delta_1 A_{bulk}=A r_{max}
\ee
where $r_{max}$ is the IR cutoff in the bulk which should be identified with a UV cutoff in the boundary.
This is the expected universal contribution to entanglement which arises from very short distance modes entangled across the boundary of ${\bf R}$.

Now  one would expect that as $L$ is increased the bulk surface penetrates deeper into the IR eventually entering the
scaling region eq.(\ref{metent}). For large enough $L$ one expects that the radial variable $\xi$ stays approximately constant in the UV and undertakes most of its excursion from $L$ to $0$ in this scaling region.
 We will make these  assumption here  and proceed. These assumptions can  be  verified   numerically for the interesting range
of parameters and we will comment on this further below.

With these assumptions the contribution from the scaling region for the minimal area surface $\delta_2 A_{bulk}$ can
be  estimated by
a simple scaling argument. We can neglect the
change in $\xi$ before the surface enters the scaling region and take
 its value  at the boundary of the scaling region\footnote{By the boundary of the scaling region we mean the region where the metric begins to significantly depart from eq.(\ref{metent}). This happens as we go to larger values of $r$;
for even larger values the metric becomes
AdS space.} which we denote as $\xi_0$ to be $L$.
Under the scaling symmetry eq.(\ref{rscale})-eq.(\ref{xyscale}),  which takes
\be
\label{scen}
r\rightarrow \lambda r, \xi \rightarrow \lambda^{1-\beta-\gamma} \xi,
\ee
we see from eq.(\ref{acs}) that
\be
\label{scbua}
\delta_2 A_{bulk} \rightarrow  \lambda^{2(1-\gamma)} \delta_2 A_{bulk}
\ee

Now by choosing $\lambda$ in eq.(\ref{scen}) to be
\be
\label{vallambdae}
\lambda=L^{1\over \gamma+\beta-1}
\ee
we can set the rescaled value for $\xi_0$ to be unity.
In terms of the rescaled variable the minimisation problem has no scale left and $\delta_2 A_{bulk}$ must be order unity.
This tells us that when $\xi_0=L$
\be
\label{finalvala}
\delta_2 A_{bulk}\sim L  L^{\gamma-\beta -1 \over \gamma+\beta-1}.
\ee
Note also that with $\xi_0$ set equal to unity the surface would  reach a minimum  value  at a radial value of $r_{min}$
which is
 of order unity. Thus before  the rescaling
\be
\label{bres}
r_{min}\sim \lambda=L^{-({1\over \beta+\gamma-1})}.
\ee

Now we are ready to consider  different regions in the $(\gamma,\beta)$ parameter space.  From eq.(\ref{case1})
  we see that $\beta>0$ and from eq.(\ref{condgamma})  that $\gamma>1/2$.
\begin{itemize}
\item  $\gamma>1+\beta$: In this case  we see from eq.(\ref{finalvala})  and (\ref{cba}) that
 $\delta_2 A_{bulk} > \delta_1 A_{bulk}$ for sufficiently big $L$ and fixed UV cutoff $r_0$.
Thus the dominant contribution to the area for sufficiently big $L$  comes
the scaling region.  The entanglement is then given by
\be
\label{entaa}
S_{EE}\sim N^2 (L\mu) (L \mu )^{\gamma-\beta-1\over \gamma+\beta-1}
\ee
with an additional term proportional to $L$ in units of  the UV cutoff $r_0$.
In eq.(\ref{entaa})  we have introduced the scale $\mu$ to make up the dimensions.
We remind the reader that this scale stands for  the  chemical potential which is the only  length scale  in the boundary theory.
We see from eq.(\ref{entaa}) that the entanglement  grows with $L$ with a power faster than unity.
Also notice that from eq.(\ref{bres})  $r_{min}$ decreases with increasing $L$ in accord with our expectation that the
 surface penetrates further into the scaling region as $L$ is increased.
\item A special case of importance is when $1+\beta=\gamma$. Here the term $(L\mu)^{\gamma-\beta-1\over \gamma+\beta-1}$ is
replaced by a $\log$,
\cite{Ogawa},  resulting in eq.(\ref{entaa}) being replaced by
\be
\label{entac}
S_{EE}\sim N^2 (L\mu) \log(L\mu)
\ee
\item  $1+\beta>\gamma>1-\beta$ : In this case we see  from eq.(\ref{finalvala}) and eq.(\ref{cba})
that the contribution $\delta_2 A_{bulk}$ grows with $L$ with
a power less than unity and therefore  the contribution made by the scaling region to the total area is less significant than
$\delta_1 A$ which is linear in \footnote{ We should note that the scaling argument tells us  that  $r_{min}$ does decrease with
increasing $L$, eq.(\ref{bres}), so that the surface does get further into the scaling region as $L$ increases.}$L$.
In this region of parameter space the entanglement entropy is therefore
 dominated by the short distance contributions and  given by
\be
\label{entab}
S_{EE}\sim N^2 {L\over a}
\ee
where $a = {1\over r_{max}}$ is an UV cutoff in the system.
\item $\gamma<1-\beta$:  In this case $r_{min}$ does not decreases with increasing $L$, actually the surface stops entering the
scaling region and our considerations based on the  scaling symmetry are not relevant. The entanglement entropy is again given by
 eq.(\ref{entab}).
\end{itemize}

 One can calculate  the minimal area surface numerically for cases
when $\gamma>1+\beta$ and also $1+\beta>\gamma>1-\beta$.
This gives  agreement with the above discussion including the scaling behaviour for $r_{min}$ eq.(\ref{bres})
 in these regions of parameter space.

Let us make some more  comments now.
In the case where the near horizon geometry is of extreme RN type, i.e., $AdS_2\times R^2$ we have $\beta=0, \gamma=1$.
This case needs to be dealt with separately. Here the entanglement entropy scales like the volume and equals the Beckenstein-Hawking entropy of the corresponding region in the boundary theory.
\be
\label{entd}
S_{EE} \sim N^2 (L\mu)^2
\ee

The discussion in \S2 was in terms of the parameters $(\alpha,\delta)$ while  here we have used $(\beta,\gamma)$.
The relation between these parameters is obtained from   eq.(\ref{case11}) and summarised in Table 1.  We see that $\gamma>1+\beta$
corresponds to Region D. The line  $\gamma=1+\beta$ corresponds to $\alpha=-3\delta$ at the interface between D and E.
The condition $1+\beta>\gamma>1-\beta$ corresponds to A and E and finally $\gamma<1-\beta$ to B and C.

\subsection{Entanglement with a Small Magnetic Field}
\label{enmag}
We are now ready to consider the effects of a small magnetic field.
The behaviour of the gravity solutions was discussed in \S2 where it was shown that  the analysis breaks up into various cases.
 In all cases we will take the solution to approach AdS space in the UV. The resulting behaviour
of the solution was discussed for the various cases in subsection \ref{addcom}.
For $r\gg  \mu$ the solution is $AdS$ space while   for $r_* \ll r \ll  \mu$ it is of
electric scaling type ($r_*$ is defined in eq.(\ref{valrs})). What happens for $r\ll r_*$ depends on the various cases.

\begin{itemize}
\item {Case II.} $|\alpha|>\delta$: In this case the geometry for
 $r\ll r_*$  is $AdS_2\times R^2$.
Let us start with a boundary circle of very small radius and slowly increase its size.
When the  radius $L \mu  \ll 1$ the entanglement entropy is given by eq.(\ref{entab}).
When $L\mu  \sim 1$ the surface begins to penetrate the electric scaling region and as
$L$ increases we see from  eq.(\ref{bres}) that $r_{min}$ decreases.
When $r_{min}$ reaches $r_*$ the surface begins to enter the region where the magnetic field
has an appreciable effect on the geometry. Using eq.(\ref{bres}) this corresponds to
\be
\label{pel}
L \sim L_*  = {1\over \mu} \left({Q_e^2 \over Q_m^2}\right)^{-{1\over (4 \alpha k)(\beta+\gamma-1)}}.
\ee
For
\be
\label{condlstar}
L_*\gg L\gg {1\over \mu}
\ee
 the entanglement entropy is given by the calculation in the electric
solution. Thus for $-3\delta <\alpha<-\delta$ it  grows faster than $L$
with  an additional fractional power while   for $\alpha=-3\delta$ it is  logarithmically enhanced.
For other values of $(\alpha,\delta)$ which lie in this region  the entanglement entropy is proportional to $L$
and is   dominated by the UV contribution.

Finally, when $L \gg L_*$  the surface enters into the near-horizon $AdS_2\times R^2$ geometry.
Now the entanglement entropy  grows like $L^2$ and is given by
\be
\label{entacase2}
S \sim L^2 N^2 \mu^2 \left({Q_m \over \mu^2}\right)^{\alpha+\delta \over 2 \alpha}
\ee
This is  an expression analogous to the Beckenstein-Hawking
entropy, eq.(\ref{entern}), but with $L^2$ now being the volume of the region of interest.
For the case of special interest, $\alpha=-3\delta$, this becomes
\be
\label{csp}
S\sim L^2 N^2 \mu^2 \left({Q_m\over \mu^2}\right)^{1\over 3}.
\ee

\item {Case III.} $0<\alpha<\delta$: In this case the magnetic field is important at small $r$.
For $L_*\gg L\gg {1\over \mu}$ the entanglement is given by the electric theory, it
 goes like eq.(\ref{entab}) and arises
dominantly
 due to short distance correlations.  The geometry in the deep IR could be
 the  magnetic scaling solution discussed in subsection \ref{case3}. If this is correct for $L\gg L_*$
the entanglement will continue to go like eq.(\ref{entab}).

\item {Case I.} $-\delta<\alpha<0$: In this case the magnetic field is not important in the IR and the
 the entanglement entropy is given by eq.(\ref{entab}) both when $L<{1\over \mu}$ and $L>{1\over \mu}$.

\end{itemize}
\section{Concluding Comments}
In this paper we have studied a system of gravity coupled to an Abelian gauge field and a dilaton.
This system is of  interest from the point of view of studying fermionic matter at non-zero charge density.
Some of our   key results were  summarised in  \S1.
We end in this section with some concluding comments.

\begin{itemize}
\item
For the case $|\alpha|>\delta$ (Case II in our terminology) we saw that the magnetic field is a relevant perturbation in the IR
and the inclusion of a small magnetic field
changes the behaviour  significantly making the zero temperature  thermodynamic entropy  extensive and the entanglement grow like
the volume. In particular this happens  along the line $\alpha=-3\delta$ ($\theta=1$, eq.(\ref{defztheta})),
where the electric theory has an entanglement entropy
of the form
eq.(\ref{EEfermi}) suggesting the presence of a Fermi surface. 

\item
It is well known that  an extensive ground state  entropy can arise in the presence of a magnetic field
due to   partially filled Landau levels.
When this happens in a free fermionic theory   the entropy  scales like $Q_m$ while,  in contrast, for  the dilaton system the  dependence on the
the magnetic field  is typically more exotic, eq.(\ref{entacase2}).
E.g., with   $\alpha=-3\delta$
the entropy goes like $Q_m^{1/3}$, eq(\ref{csp}). Such a   non-trivial  exponent
 suggests that the ground state is more interesting and strange.

\item
A notable feature about how the entanglement entropy behaves in all the  cases we have studied is that it never decreases in the
IR, i.e., as one goes to regions of larger and larger size ($L$)in the boundary.  For instance, consider the case
where $-3 \delta \le \alpha <-\delta$.
In this case for very small $L$ is it given by eq.(\ref{entab}) and dominated by short distance correlations of the CFT.
At intermediate values of $L$, meeting the condition eq.(\ref{condlstar}), it goes like eq.(\ref{entaa})
 and is enhanced compared to   the $L$
dependence by an additional fractional power of $L$ or a $\log(L)$, eq.(\ref{entac}).  Finally at very large values of $L$  it grows like
the volume $L^2$, eq.(\ref{entd}). We see that as $L$ increases the entanglement increases monotonically \footnote{Our scaling argument does not directly fix the sign of the entanglement entropy in eq.(\ref{entaa}) and eq.(\ref{entac}).
However it is clear that the
sign must be positive since the corresponding  contribution to the surface area
 is bigger than the contribution from  the UV for fixed $r_{max}$ as
$L\rightarrow \infty$, and the total surface area must be positive.} .
In other cases while the detailed behaviour is different this feature is still true.
 These observations are in agreement with \cite{LM} where a renormalised version of the entanglement entropy was defined and
it was suggested that in $2+1$ dimensions this entropy would monotonically increase. It is easy to see
that the behaviour of the entanglement entropy we have found implies that
the renormalised entanglement
entropy of \cite{LM} is  monotonic and increasing.

\item
In \cite{IKNT}  the behavior of a probe fermion in the bulk in the electric hyperscaling violating geometry was discussed.
This corresponds to calculating the two point function of a gauge invariant fermionic correlator in the boundary.
It is notable that the region in  parameter  space
 where Fermi liquid behaviour was found to occur for this correlator is exactly the region $|\alpha|>\delta$
for which we have found that  the geometry flows to an  $AdS_2\times R^2$ endpoint in the IR \footnote{
The remaining region $|\alpha|<\delta$ does not exhibit Fermi liquid behaviour, this includes Case I of \S2
 where the magnetic field
is irrelevant and also Case III of \S2  where it is relevant but where
we have not identified a definite IR end point to which the solution flows.}.
It would be worth  understanding  this seeming coincidence more deeply.
It is also worth mentioning that in \cite{IKNT}
marginal Fermi liquid behaviour was found   when $|\alpha|=\delta$.
This region lies at the boundary of the region $|\alpha|>\delta$ where an $AdS_2\times R^2$ endpoint arise \footnote{In fact the nature of the attractor changes at this boundary. E.g. in the purely electric case
when $\alpha=-\delta$ the dilaton is a flat direction of the attractor potential  and not fixed to a unique value in the IR,
similarly for
$\alpha=\delta$ in the purely magnetic case.}.

\item
Our focus in this paper was on taking an electrically charged system and including a small magnetic field.
However, it is worth pointing out that the magnetic solutions with no electric field present ($Q_e=0$)
 are also of considerable interest in their own right. These solutions can be obtained by taking
\be
\label{emtrans}
Q_e\rightarrow Q_m
\ee
 and $\alpha\rightarrow -\alpha$ in the
solutions eq.(\ref{case1}), eq.(\ref{case11}).
  For a choice of parameters, which now meet the condition $\alpha=3 \delta$,
 the resulting entanglement entropy has the form eq.(\ref{EEfermi}) which suggests the presence of a Fermi surface
even though the charge density is now vanishing. It would be worth understanding the resulting state better in the field theory.
The transformation eq.(\ref{emtrans}) is an electromagnetic duality transformation, and should act by exchanging charged particles
 with vortices in the field theory\cite{Witten:2003ya},\cite{Hartnoll:2007ih}. These vortices perhaps form a Fermi surface resulting in the logarithmic enhancement of the entropy.

\item
We have not included an axion field in our analysis. Such a field is natural to include once  the dilaton is present and it
can have important consequences once the magnetic field is also turned on as was discussed in \cite{GIKPTW}.
For example it was shown in \cite{GIKPTW} for the case $\delta=0$ that  in the presence of the  axion the entropy at extremality continues to
  vanishes in the presence of a magnetic field.  Once the axion is  included  we need to allow for the potential to also depend
 on it, this leads to considerable choice in the kinds of models one can construct.
To remove some of this arbitrariness it would be worth including the axion within the context of models which
arise in string theory or at least gauged supergravity.

\item
More generally, the time seems now ripe to systematically embed models of this type in string theory and supergravity.
Some  papers in this direction have already appeared \cite{Dong:2012se}, \cite{Narayan:2012hk},\cite{Singh:2012un},\cite{Dey:2012tg},
\cite{Dey:2012rs},\cite{Perlmutter:2012he}.
  It would be worth understanding these constructions better and
 also gaining a better understanding of  their dual field theory descriptions.

\end{itemize}

\bigskip
\centerline{\bf{Acknowledgements}}
\medskip
We are grateful to  K. Damle, S. Kachru, G. Mandal, S. Minwalla, R. Sensarma, T. Takayanagi and V. Tripathy for discussion.
SPT thanks the organisers of the Strings Discussion meeting at the ICTS, Bangalore, for a stimulating conference and
 acknowledges the support of a J.C. Bose Fellowship from the Dept. of Science and Technology,
Govt. of India.  We all
 acknowledge the support of DAE, Government of India and most of all are grateful to the people of India
for generously supporting research in string theory.

\newpage

\appendix
\section{Appendix : Numerical Interpolation.}
In this appendix we consider Case II solutions which were discussed in section \ref{subsectioncase2}
and establish that the deep IR solution  is indeed $AdS_2\times R^2$.
We establish this  numerically by integrating outwards from the $AdS_2\times R^2$  near horizon
solution discussed in subsection \ref{subsectioncase2}
  and showing that the system   approaches the electric scaling solution. For a suitable potential
we find that the electric scaling solution in turn  finally asymptotes to $AdS_4$ in the UV.
Our numerical work is done using the Mathematica package.

We divide the discussion into three parts. In the first part we identify two
perturbations in the $AdS_2\times R^2$ region  that grow towards the UV.
In the second part, by choosing an appropriate combination of these two
perturbation, we numerically integrate outwards taking the scalar potential to
be $V(\phi)= -|V_0| e^{2\delta \phi}$, eq.(\ref{defV}). At moderately large radial distances we get
the electric scaling solution. In the last part, taking the potential to be of
the form $V(\phi)= -2 |V_0| \cosh(2\delta \phi)$ \footnote{Note, for this last part we present results for  the case where
 $\alpha>\delta$ so that in the electric scaling solution $\phi\rightarrow \infty$ in the deep IR.
 As a result  the modified potential $-2 |V_0| \cosh(2\delta \phi)$ can
 be approximated by $-|V_0| e^{2\delta \phi}$ in the IR. In the UV we then find that the modified potential
 allows for an $AdS_4$ solution. A similar analysis can be done when $\alpha<-\delta$, using a suitably modified potential.}, we continue the numerical integration towards larger $r$ and
show that the geometry asymptotes to $AdS_4$.

\subsection{The perturbations}

To identify the perturbations in the $AdS_2\times R^2$ solution discussed in \ref{subsectioncase2}
we consider perturbations of this solution
 of the following
form for the metric components and the dilaton,
\begin{eqnarray}
 a(r) &=& C_a r \left[1 +a_{c1} r^\nu +{\cal O}(r^{2\nu}) \right]~\label{eqA1}\\
 b(r) &=& b_0 \left[1 + b_{c1} r^\nu+{\cal O}(r^{2\nu})\right]~\label{eqA2}\\
 \phi(r) &=& \phi_0 + \log\left[1 +\phi_{c1} r^\nu+{\cal O}(r^{2\nu}) \right].~\label{eqA3}
\end{eqnarray}
Note that this is a perturbation series in $r^{\nu}$ which is valid
in the near horizon region where $r \ll 1$. Equations (\ref{em1}-\ref{em4}) can be solved
to
leading order in $r^{\nu}$ to find perturbations which are relevant towards the  UV,
i.e with  $\nu>0$. We find two such perturbations parametrized by their
strengths $\phi^{(1)}_{c1},a^{(2)}_{c1}$ which are given below :
\begin{eqnarray}~\label{pertsol1}
a^{(1)}_{c1} &=& {2 \delta  \over
1+\alpha^2-\delta^2+\sqrt{1+4\alpha^2-4\delta^2}}\phi^{(1)}_{c1}\\
b^{(1)}_{c1} &=& 0\\
\nu_1 &=& {1\over2}(-1+\sqrt{1+4\alpha^2-4\delta^2})
\end{eqnarray}
and
\begin{eqnarray}
b^{(2)}_{c1} &=& -{3(-2+\alpha^2-\delta^2)\over
2(-2+\alpha^2-2\delta^2)}a^{(2)}_{c1}\\
\phi^{(2)}_{c1} &=&- {3\delta \over (-2+\alpha^2-2\delta^2)}a^{(2)}_{c1}\\
\nu_2 &=& 1
\end{eqnarray}
Note that $b^{(1)}_{c1}$ vanishes in the first of the above perturbations, as a result the first correction to $b(r)$ starts
 at second order. Actually, we found it  important  to go to second order in the first of the above perturbations for carrying
out the numerical integration satisfactorily. We will not provide the detailed expressions for these second order corrections here since they are cumbersome.

\subsection{Scaling symmetries}
The system of equations eq.(\ref{em1})-eq.(\ref{em4})  has two scaling symmetries.
\begin{eqnarray}
a^2\rightarrow \lambda_1 a^2, & & r^2 \rightarrow \lambda_1 r^2 \label{symone} \\
(Q_e^2, Q_m^2)  \rightarrow \lambda_2 (Q_e^2,Q_m^2), & & b^2 \rightarrow \lambda_2 b^2 \label{symtwo}
\end{eqnarray}
The first  scaling symmetries  can be used to set    \footnote{We cannot change the sign of $a^{(2)}_{c1}$ by using
the symmetries. The above sign is necessary for the solution to flow to the electric scaling solution in the UV.}
$a^{(2)}_{c1}=-1$. The second scaling symmetry can then be used to set $Q_e=1$.
In addition we can choose units so that $|V_0|=1$.
With these choices the system of equations has two parameters, $\phi^{(1)}_{c1}$, which characterises the first perturbation, eqn. (\ref{pertsol1}),
and $Q_m$.
 Numerical analysis shows that for a given choice of $Q_m\ll \mu^2$,  $\phi^{(1)}_{c1}$  needs to be tuned very precisely to
ensure that the solution flows to the electric scaling
solution. Otherwise, for example with the modified potential considered in subsection \ref{Asubsection3} below, the  solution can  flow from the
$AdS_2\times R^2$ region in the IR directly to $AdS_4$, as $r\rightarrow \infty$,
 without passing close to the electric scaling solution at intermediate values of $r$.

\subsection{Numerics : $AdS_2\times R^2$  to the Electric Scaling Solution}
We will illustrate the fact that
  the solution  evolves from the $AdS_2\times R^2$ geometry to the electric scaling solution once the
values of $\phi^{(1)}_{c1}$ is suitably chosen with one example here. Similar behaviour is found for other values of
$(\alpha,\delta)$, which lie in Case II \footnote{The example we choose here  has $\alpha,\delta>0$.
Similar results are also obtained
when $\alpha<0,\delta>0$.}

The example we present here has  $\alpha=1, \delta= 0.6$ (satisfying $|\alpha|>\delta$).
We will present the data here for the case when $Q_m=10^{-4}$ the behaviour for other values of $Q_m\ll \mu^2$ is similar.
 It turns out that in this case we
have to fine tune the value of this parameter to  be near  $\phi^{(1)}_{c1}=-0.3173$ so as
to obtain an  electric scaling solution at intermediate $r$.

\begin{figure}[h]
\begin{center}
 \includegraphics[scale=0.8]{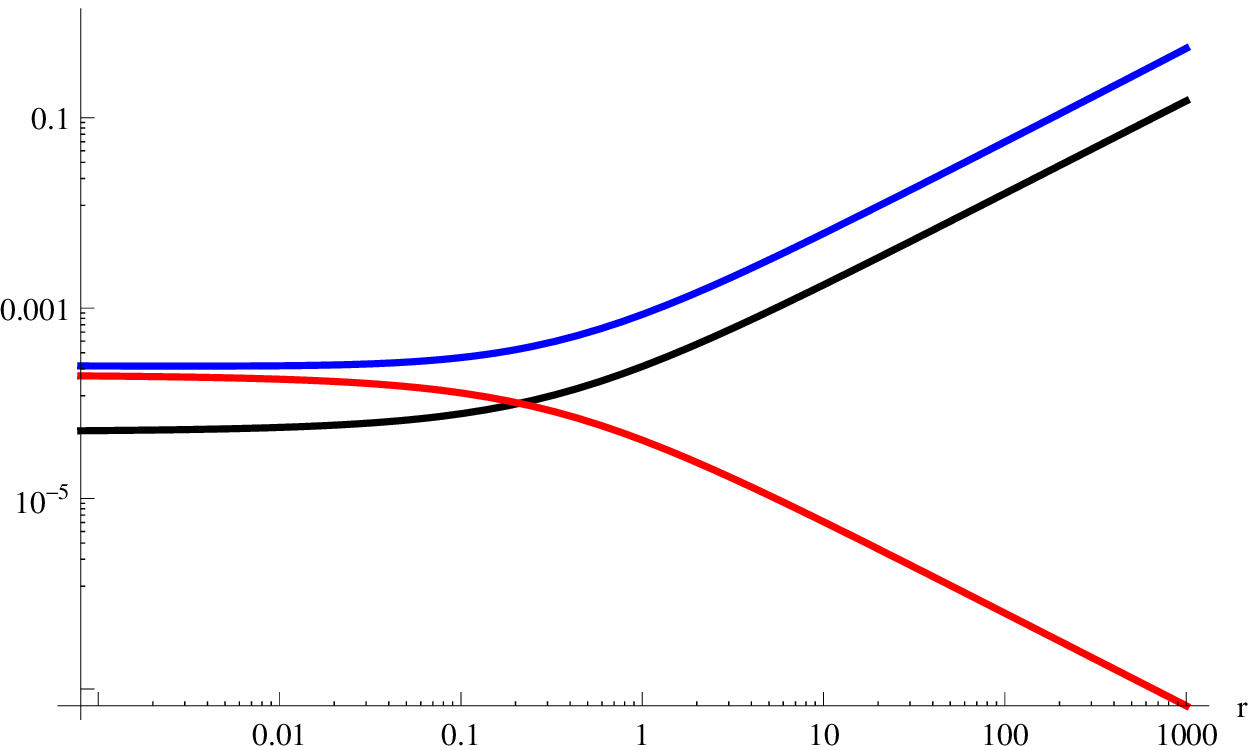}
\caption{Different contributions to $V_{eff}$ in Log-Log plot. Blue$\rightarrow$
Scalar Potential $b^4(r) {e^{2\delta \phi}\over 2}$, Black$\rightarrow$ $Q_e^2
e^{-2\alpha \phi}$ term, Red$\rightarrow$ $Q_m^2 e^{2\alpha \phi}$ term.}
\label{veffloglog1}
\end{center}
\end{figure}

Evidence for the  electric scaling solution  can be obtained by examining  the relative contributions
that various terms make in the effective potential eq.(\ref{Veff}). In the electric scaling region the
contribution that the $Q_m^2$ dependent term makes must be smaller than the
$Q_e^2$ dependent term and the scalar potential which in turn  must scale in the same way.
 Fig.(\ref{veffloglog1}) shows
the different contributions to $V_{eff}$ made by  the terms, $e^{-2\alpha \phi}
Q_e ^2, e^{2\alpha \phi} Q_m ^2$ and $b^4(r) {e^{2 \delta \phi} \over 2}$  in a Log-Log plot. Clearly the $Q_e^2$ term is growing as the same
power of $r$ as the scalar potential $e^{2 \delta \phi}$ term and $Q_m ^2$ is
 subdominant.

\begin{figure}[h]
\begin{center}
 \includegraphics[scale=0.69]{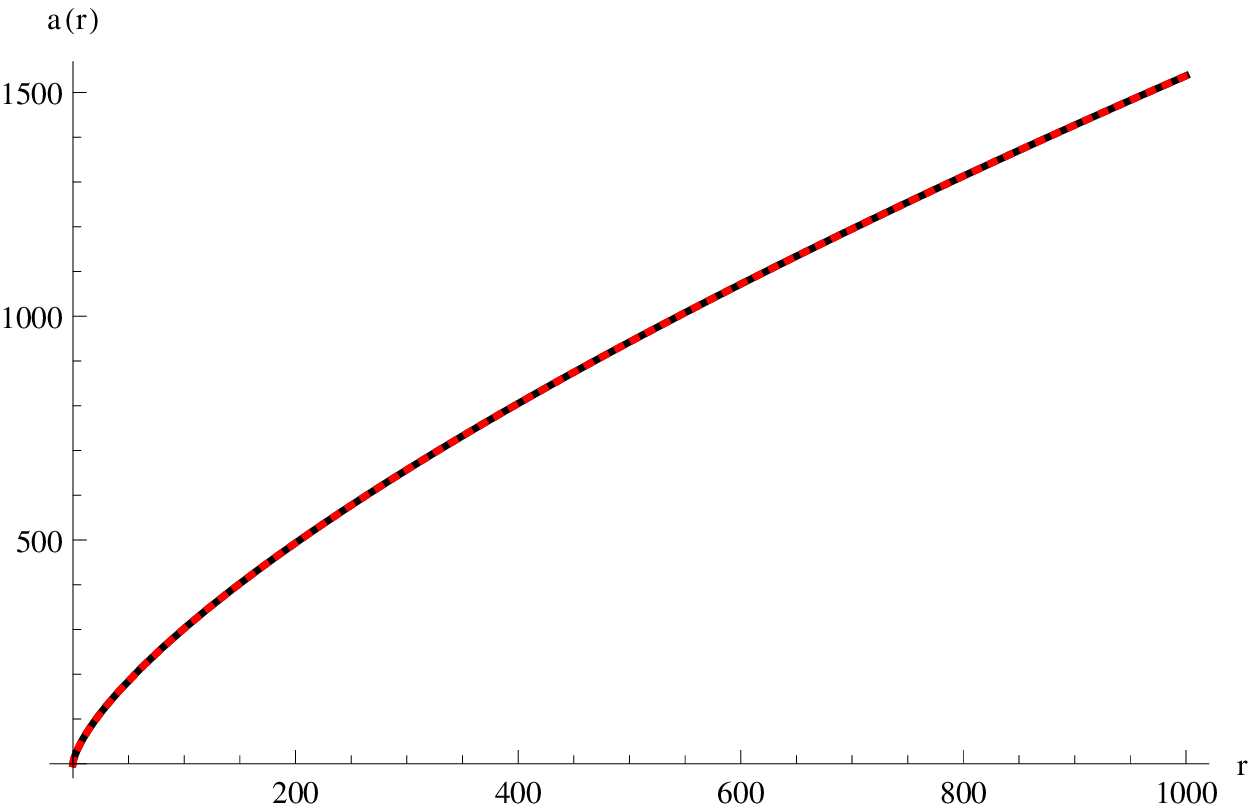}
 \includegraphics[scale=0.69]{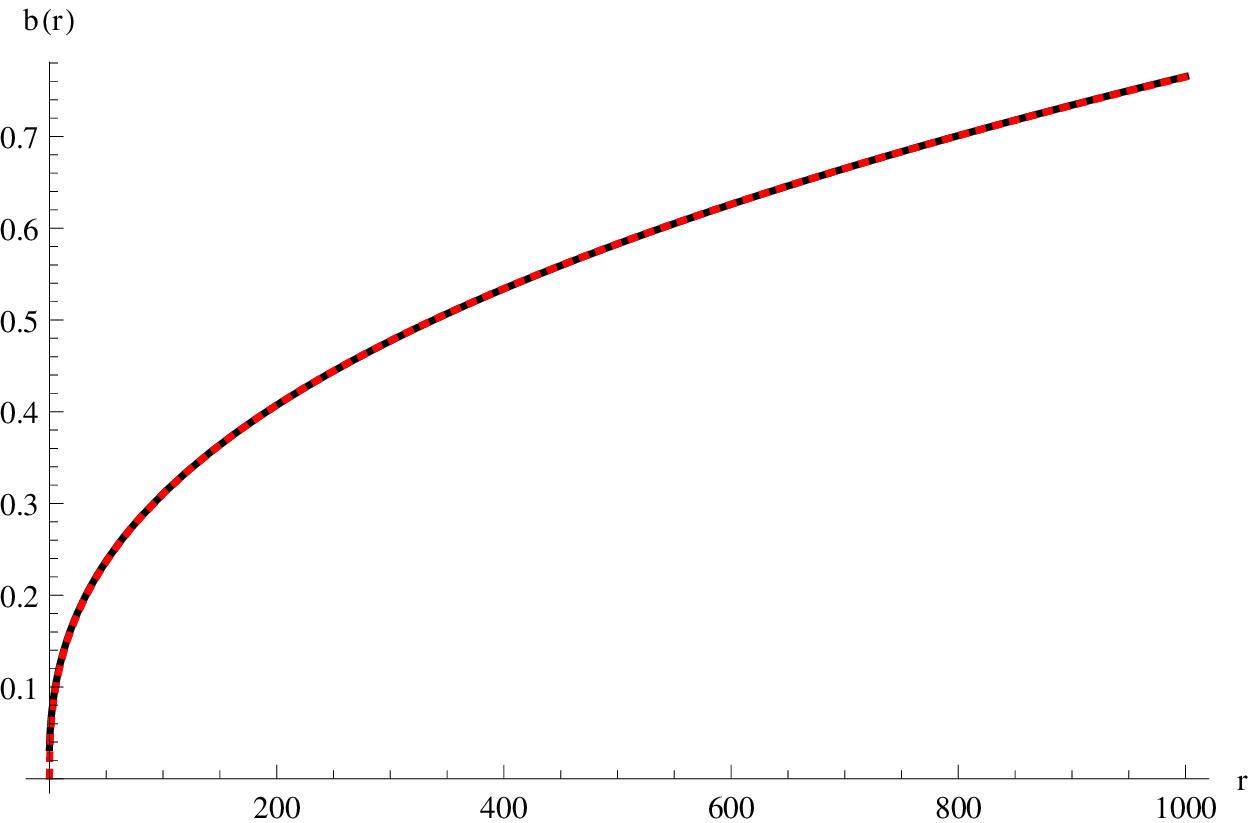}
\caption{In the top plot, $a(r) \sim r^{\gamma}$ with $\gamma_{Fit}=0.706$
whereas $\gamma_{Analytical}=0.707$. In the bottom plot, $b(r) \sim r^{\beta}$
with $\beta_{Fit}=0.391$ whereas $\beta_{Analytical}=0.390$.} \label{abvsr}
\end{center}
\end{figure}

\begin{figure}[h]
\begin{center}
 \includegraphics[scale=0.69]{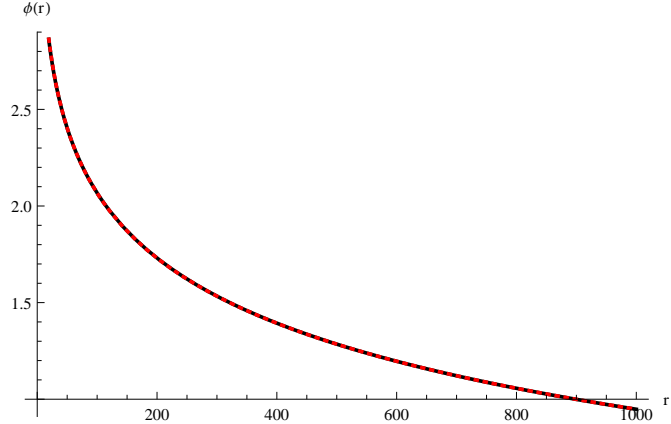}
\caption{$\phi=k \ log[r]$. $k_{Fit}= -0.4858$ as compared to
$k_{Analytical}=-0.4878$.} \label{phivsr1}
\end{center}
\end{figure}

 Fig.(\ref{abvsr}) and Fig.(\ref{phivsr1})
show the plots of metric components $a(r),b(r)$ and the scalar $\phi(r)$
obtained numerically. Each of them is fitted to a form given in eq(\ref{case1}).
 We see that the fitted parameters agree well with the analytic values for  $\beta,\gamma$ and
$k$  obtained from  eq.(\ref{case11}) with  $(\alpha,\delta)=(1,0.6)$. This confirms that the system  flows to
the electric scaling solution.
\subsection{Numerics : $AdS_2\times R^2$ $\rightarrow$ Electric Scaling $\rightarrow$   $AdS_4$.}~\label{Asubsection3}
Here we show that on suitably modifying the potential so that the IR behaviour is essentially left unchanged the solution
which evolves from the $AdS_2\times R^2$ geometry in the deep IR to the electric scaling solution can be further extended to
become asymptotic $AdS_4$ in the far UV.

We will illustrate this for the choice made in the previous subsection: $(\alpha,\delta)=(1,0.6)$, $(Q_e=1, Q_m=10^{-4})$.
For this choice of $(\alpha,\delta)$ it is easy to see from eq.(\ref{case1}), eq.(\ref{case11}) that $\phi \rightarrow \infty$
in the IR of the electric scaling solution, and from eq.(\ref{atphi}) that it continues to be big in the $AdS_2 \times R^2$ geometry
once the effects of the magnetic field are incorporated. We will modify the potential to be
\be
\label{modphi}
 V(\phi)=-2 |V_0| \cosh(2\delta \phi)
\ee
instead of eq.(\ref{defV}). For $\phi\rightarrow \infty$ we see that this makes a  small change, thus our analysis
in the previous subsection showing that the solution evolves from the $AdS_2 \times R^2$ geometry to the electric scaling solution
will be essentially unchanged. However, going to larger values of $r$ the modification in the potential will become important.
This modified  potential has a maximum  for the dilaton at $\phi=0$  and   a corresponding
  $AdS_4$ solution with \footnote{This example was analysed in \cite{IKNT}.
The dilaton lies above the BF bound of the resulting $AdS_4$
theory for our choice of parameters.} $R_{AdS}^2 = -{3 \over V_0}$.
We find by numerically integrating from the IR  that   the solution evolves  to this $AdS_4$ geometry in the UV.

\begin{figure}[h]
\begin{center}
 \includegraphics[scale=0.8]{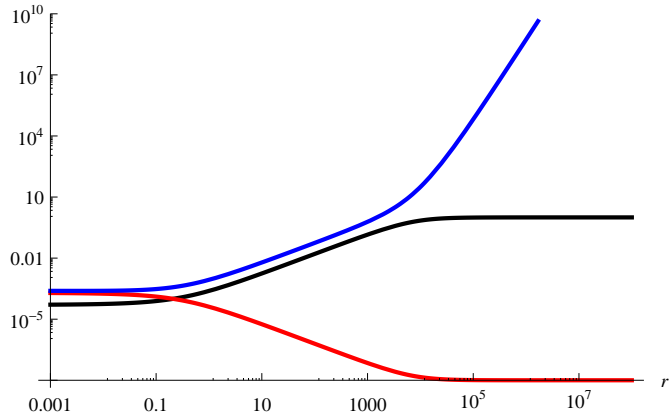}
\caption{Different contributions to $V_{eff}$ in Log-Log plot. Blue$\rightarrow$
Scalar Potential $b^4(r) \cosh(2 \delta \phi)$, Black$\rightarrow$ $Q_e^2
e^{-2\alpha \phi}$ term, Red$\rightarrow$ $Q_m^2 e^{2\alpha \phi}$ term.}
\label{veffloglog2}
\end{center}
\end{figure}

To see this first consider a plot of the three different contributions to $V_{eff}$
 proportional to $e^{-2\alpha \phi} Q_e ^2, e^{2\alpha \phi} Q_m ^2$
and $b^4(r) \cosh(2 \delta \phi)$   shown in   Fig(\ref{veffloglog2}).
We see that there are three distinct regions. In the far IR $AdS_2\times R^2$ region,
 the three contributions are comparable. At intermediate $r$ where we expect an electric scaling solution on the basis of the discussion of the previous subsection the magnetic field makes a subdominant contribution and the other two contributions indeed
 scale in the same way.  Finally at very large $r$, in the far UV,
 the cosmological constant is dominant as expected for an $AdS_4$
solution.

\begin{figure}[h]
\begin{center}
 \includegraphics[scale=0.69]{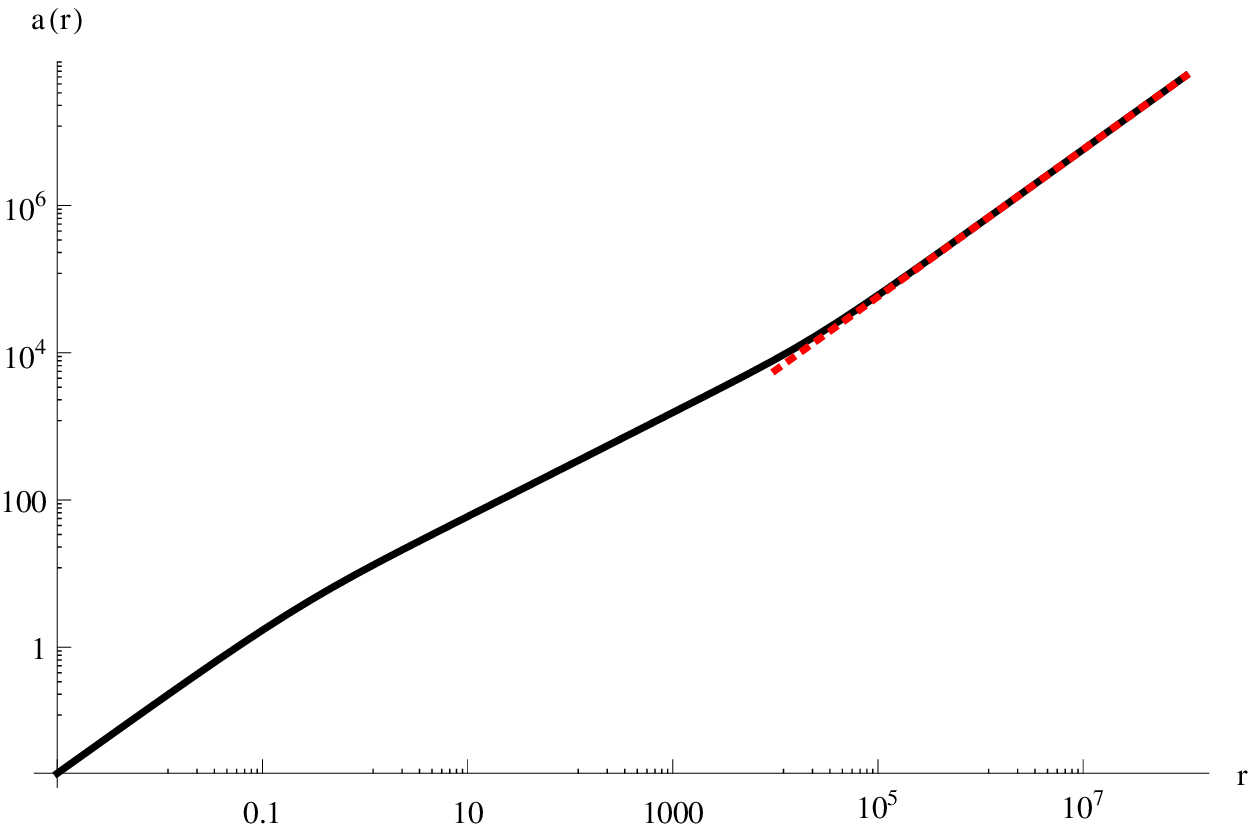}
 \includegraphics[scale=0.69]{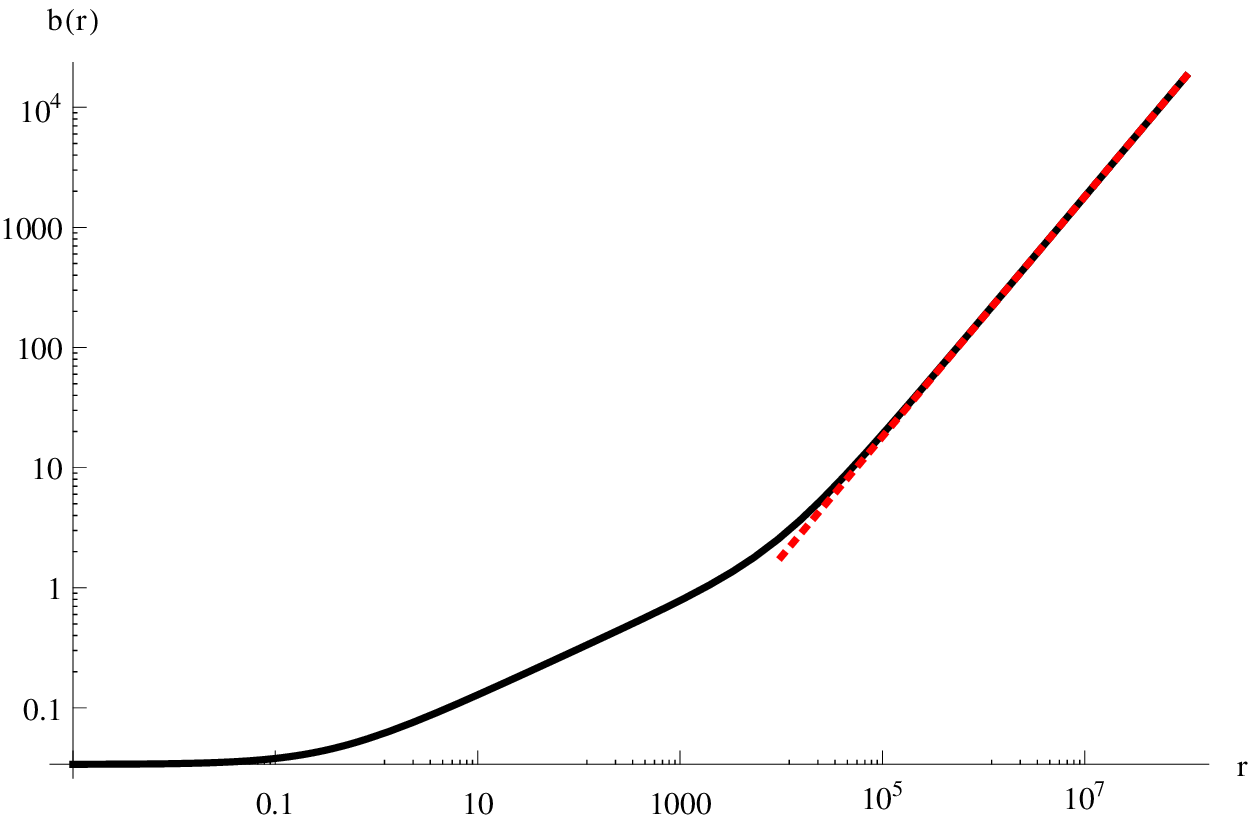}
\caption{In the Log-Log plot both  curves show three different slopes for the
metric components, which correspond to the three regions $AdS_2\times R^2$, Electric Scaling Region and
 $AdS_4$ respectively.}
\label{logabvsr}
\end{center}
\end{figure}

\begin{figure}[h]
\begin{center}
\includegraphics[scale=0.69]{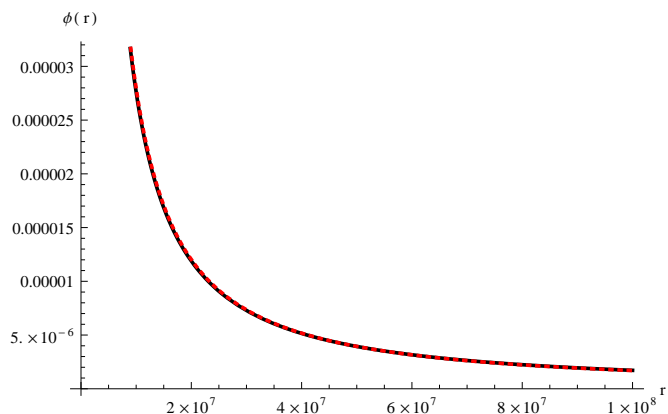}
\caption{$\phi(r)$ approaches zero like $r^{-1.22}$ as $r\rightarrow \infty$.
This agrees with the fall off of non-normalizable component of dilaton in
$AdS_4$.} \label{phibvsr2}
\end{center}
\end{figure}

 We also show the metric components $a(r), b(r)$ in a Log-Log plot in
 Fig(\ref{logabvsr}) and (\ref{phibvsr2}). Once again, we can see three distinct slopes for  $a,b,$
corresponding to three different regions in the solution. In the $AdS_4$ region, as $r\rightarrow \infty$,
numerically fitting the behaviour  gives $a(r), b(r) \sim r^{0.99}$ which is in good agreement\footnote{The fit was done
for $r\sim 10^7$.} with the expected linear behaviour.
  Finally,
Fig(\ref{phibvsr2}) shows the scalar
function $\phi(r)$ settling to zero with the expected fall-off as $r\rightarrow\infty$. These
 results confirm that the system evolves to $AdS_4$ in the far UV.


\end{document}